\documentclass[aps,prd,preprint,superscriptaddress,amsmath,amssymb,showpacs]{revtex4-1}
\usepackage{dcolumn}
\usepackage{graphicx}
\usepackage{float}
\usepackage{physics}
\usepackage[colorlinks=true,allcolors=blue]{hyperref}
\begin{document}
	\title{ Chiral Condensation and Chiral Phase Diagram under Combined Rotation and Chemical Potential in Holographic QCD}
	
	\author{De-Yang Long}
	\affiliation{College of Mathematics and Physics, China Three Gorges University, Yichang 443002, China}
	
	\author{Sheng-Qin Feng}
	\email{Corresponding author: fengsq@ctgu.edu.cn}
	\affiliation{College of Mathematics and Physics, China Three Gorges University, Yichang 443002, China}
	\affiliation{Center for Astronomy and Space Sciences and Institute of Modern Physics, China Three Gorges University, Yichang 443002, China}
	\affiliation{Key Laboratory of Quark and Lepton Physics (MOE) and Institute of Particle Physics,\\
		Central China Normal University, Wuhan 430079, China}
	
	\date{\today}

\begin{abstract}
We investigate the combined effects of rotation and finite quark chemical potential on inhomogeneous chiral condensation and the chiral phase diagram within the soft-wall holographic QCD model. Using the five-dimensional AdS-RN metric, we study the spatial profile of the chiral condensate and the resulting $T - \Omega$ phase diagram under Neumann and Dirichlet boundary conditions. Increasing $\Omega$ induces strong spatial inhomogeneity: the condensate is suppressed more strongly near the edge than at the center, and vanishes at the boundary when $\Omega$ exceeds a critical value. The chemical potential $\mu$ acts as a global suppression factor, reducing the condensate magnitude without altering the spatial pattern. The $T - \Omega$ phase diagrams are investigated for different chemical potentials. For the case $\mu$ = 0, they are also studied at different distances from the rotation axis. It is found that both $\Omega$ and $\mu$ lower the chiral phase transition temperature, and their suppression effects are additive. In a rotating system, the critical temperature becomes position-dependent, decreasing with increasing distance from the rotation axis. These findings reveal a rich, spatially dependent phase structure in rotating QCD matter, relevant for non-central heavy-ion collisions.
\end{abstract}

\maketitle

\section{Introduction}\label{sec:Intro}

Quantum Chromodynamics (QCD), the fundamental theory of strong interactions, exhibits rich non-perturbative phenomena, including spontaneous chiral symmetry breaking and color confinement, which determine the structure and properties of hadronic matter. The behavior of these phenomena under extreme conditions -- such as high temperature, high density, strong magnetic fields, and rotation has become a central topic in nuclear and high-energy physics ~\cite{Zhao:2023pne,Rodrigues:2017cha,Zhao:2023pne,Chen:2020ath,Fang:2021ucy,Zhao:2022uxc,Chen:2024jet,Chen:2021gop,Dudal:2015wfn,Guo:2019joy,Chen:2019qoe,She:2017icp,Zhong:2014sua,Chen:2017lsf,Zhong:2014cda,Mo:2013qya,Yang:2025zcs}. Among these extreme scenarios, the rotating quark-gluon plasma (QGP) created in non-central heavy-ion collisions (HICs) provides a unique laboratory for exploring the interplay between rotational effects and chiral symmetry dynamics. In such collisions, local angular velocities can reach 0.2~GeV, and angular momentum  reaches 10\textsuperscript{4}$\hbar$ ~\cite{Chen:2015hfc,Jiang:2016wvv,Wang:2019nhd,Chernodub:2020qah,Ebihara:2016fwa,Chen:2022mhf,Wang:2018zrn,Hua:2024bwn,Wang:2024rim}.

Rotation induces an inhomogeneous chiral condensate, a phenomenon analogous to quantized vortex formation in rotating superfluids, and with profound implications for QCD phase transitions. Previous studies ~\cite{Colangelo:2011sr,Bartz:2017jku,Chelabi:2015gpc,Li:2016gfn,Chelabi:2015cwn,Braga:2025eiz,Lee:2009bya} in effective field theories and holographic frameworks confirm that rotation suppresses chiral condensation via the $\Omega\cdot J$ term, independent of boundary conditions. They also constructed the $T-\Omega$ phase diagram, which resembles the $T-\mu$ diagram. The similarity between the $\Omega\cdot J$ coupling (angular velocity with angular momentum) and the $\mu\cdot N$ coupling (chemical potential with conserved charge) suggests an intrinsic analogy between rotation and chemical potential in regulating chiral symmetry.

The quark chemical potential $\mu$ directly measures baryon density and is essential for dense QCD matter, e.g., in neutron stars or the early universe. It plays a pivotal role in shaping the QCD phase structure. It is well established that increasing $\mu$ drives the restoration of chiral symmetry, accompanied by the vanishing of the quark-antiquark condensate $\langle \overline{q}q \rangle$ ~\cite{Park:2009nb,Park:2011qq}. However, lattice QCD simulations, the most reliable ab initio method for non-perturbative QCD, suffer from the sign problem at finite $\mu$, hindering direct investigations of dense QCD matter ~\cite{deForcrand:2009zkb,Aarts:2012yal,Hands:2007by,Hanada:2012es,Greensite:2013gya}. This limitation has motivated the development of alternative non-perturbative approaches. Among these, holographic QCD (AdS/QCD), based on the Anti-de Sitter/Conformal Field Theory (AdS/CFT) correspondence, has emerged as a powerful tool for tackling strong-coupling problems ~\cite{Gubser:2008ny,Gubser:2008yx,Gursoy:2007cb,Gursoy:2007er,DeWolfe:2010he,Grefa:2021qvt,Dudal:2017max,Karch:2006pv,Li:2013oda,Li:2012ay,Domenech:2010nf,Wang:2024szr,Deng:2021kyd,Feng:2019boe,Zhu:2019ujc} and has consistently shown that rotation lowers the chiral transition temperature and induces spatial inhomogeneity ~\cite{Nadi:2019bqu,BravoGaete:2017dso,Sheykhi:2010pya,Awad:2002cz,Arefeva:2020jvo,McInnes:2018mwj,Mcinnes:2018xxz,Erices:2017izj,McInnes:2014haa,Li:2019swh,Dias:2013bwa,Keranen:2009re}. Nevertheless, the majority of these holographic studies have been carried out at zero chemical potential. As a result, the combined effects of rotation and finite baryon density, a scenario relevant to non-central heavy-ion collisions - remain largely unexplored.

The spatial profile of the inhomogeneous chiral condensate under rotation, the effects of finite size, and the dependence on Neumann vs. Dirichlet boundary conditions were systematically investigated in Ref.~\cite{Chen:2022mhf}. Key findings of the work included: (i) the suppression effect is determined by the product ${\Omega}r$; (ii) under Neumann boundary conditions, finite size exhibits catalysis at high $T$ and inverse catalysis at low $T$; (iii) under Dirichlet boundary conditions, finite size always shows inverse catalysis and can induce a transition from an inhomogeneous to a homogeneous phase at a critical radius. However, the combined effect of finite chemical potential and rotation remained largely unexplored. Our work extends Ref.~\cite{Chen:2022mhf} by systematically introducing $\mu$. We find that: (i) $\mu$ acts as a global suppressor of the condensate without altering the radial pattern induced by rotation; (ii) the $T-\Omega$ phase boundary shifts downward with increasing $\mu$; (iii) the critical angular velocity $\Omega_{c}$ for edge condensate vanishing decreases with increasing $\mu$. These results go beyond Ref.~\cite{Chen:2022mhf}, and provide a more complete phase diagram for rotating dense QGP relevant to non-central heavy-ion collisions. 

The structure of this paper is as follows: In Sec. II, we develop the holographic model for a plasma and extend it to include rotation and finite chemical potential. The spatial dependence of the chiral condensate on temperature, chemical potential, and angular velocity is presented in Sec. III.  In Sec. IV, we present and discuss the calculated results of the $T-\Omega$ phase diagrams for different chemical potentials and different radii of rotation. Finally, in Sec. V, we provide a summary and conclusions.

\section{HOLOGRAPHIC QCD THEORY UNDER ROTATION }\label{sec:2}

The soft-wall AdS/QCD model~\cite{Karch:2006pv} provides an effective holographic framework for studying chiral phase transitions and can be readily extended to incorporate rotational effects. This model successfully describes first-order, second-order, and crossover chiral phase transitions, as detailed in Refs.~\cite{Chelabi:2015gpc,Li:2016gfn,Chelabi:2015cwn}. In this framework, the complex matrix-valued scalar field \(X^{\alpha\beta}\) corresponds to the four-dimensional operator \(\langle \bar{q}^\alpha q^\beta \rangle\), where \(\alpha\) and \(\beta\) are flavor indices. To realize different types of phase transitions, one can introduce nonlinear terms into the potential and choose an appropriate form for the dilaton field \(\Phi\).

Rotation breaks spatial isotropy; therefore, in principle, both the metric and the fields should depend on the four-dimensional spacetime coordinates \(x^\mu\). However, given the complexity of fully numerical solutions in general relativity, the probe approximation is commonly adopted - neglecting the backreaction of matter fields on the background metric and focusing solely on the effects of rotation on flavor degrees of freedom. Assuming a fixed rotation axis, we adopt cylindrical coordinates for convenience. Furthermore, to incorporate the effect of chemical potential, we model the system at finite temperature and finite chemical potential using a charged black hole solution described by the five-dimensional AdS-RN metric as

\begin{equation}
	ds^2 = \frac{L^2}{z^2} \left[ -f(z) \, dt^2 + \frac{dz^2}{f(z)} + dr^2 + r^2 \, d\theta^2 + dx_3^2 \right],
	\label{eq:1}
\end{equation}
where
\begin{equation}
	f(z) = 1 - (1 + Q^2)\left(\frac{z}{z_h}\right)^4 + Q^2\left(\frac{z}{z_h}\right)^6,
	\label{eq:2}
\end{equation}
where \(z_h\) denotes the event horizon, and \(Q = q z_h^{3}\) represents the black hole charge subject to the constraint \(0 < Q^2 < 2\). The chemical potential and temperature are uniquely determined by the black hole charge and horizon location as
\begin{equation}
	\mu = \kappa \frac{Q}{z_h},
	\label{eq:3}
\end{equation}
and
\begin{equation}
	T = \frac{1}{\pi z_h} \left( 1 - \frac{Q^2}{2} \right),
	\label{eq:4}
\end{equation}
where \(\mu\) is the quark chemical potential, and \(\kappa\) is a dimensionless coefficient (we will set \(\kappa = 1\) for simplicity~\cite{Braga:2017bml,Sin:2007ze,Colangelo:2010pe,Colangelo:2012jy}).

We emphasize that the metric of Eq. (1) is the non-rotating AdS-RN background. Rotation enters the system through the boundary condition of the gauge field $A_\theta$ (which sources the angular momentum current) and through the centrifugal-like term $A_\theta^{2}/r^{2}$ in the scalar equation. This approach neglects the backreaction of rotation on the metric, which is valid when the rotational energy is small compared to the thermal energy and the inverse horizon scale, i.e.,  $\Omega \ll T$ and $\Omega \ll 1/z_{h}$. Under these conditions, the geometry remains effectively non-rotating. For the parameters considered ( $\Omega \leq 0.05$ ~GeV, $T \geq 0.1$ ~GeV), this approximation remains under control. A fully backreacted rotating metric (e.g., Kerr-AdS) would be necessary for larger ~$\Omega$, which we leave for future study.

The soft-wall AdS/QCD model has a key advantage: it lifts the global \( SU(N_f) \times SU(N_f) \) chiral symmetry of 4D space to a 5D gauge symmetry. This provides a clear correspondence between 4D operators and 5D fields. We can also extend the global symmetry to \( U(1) \times SU(N_f) \times SU(N_f) \) by introducing an additional \( U(1) \) symmetry. The conserved current of four-dimensional symmetry \( O_\mu^a = \langle \overline{q} \gamma_\mu \tau^a q \rangle \) (where \( \tau^a \) is the generator of the considered group) corresponds to the gauge field \( A^a_\mu(z,x) \) in five-dimensional space. However, these gauge fields are often neglected when only the medium background is considered. In contrast to static equilibrium systems, a rotating medium necessarily possesses nonzero currents. To simplify the analysis, we focus on the \( U(1) \) current, leading to the probe action as
\begin{equation}
	S_M = -\int d^5 x \, \sqrt{-g} \, e^{-\phi(z)} \left\{ \operatorname{Tr} \left[ \left( D^M X \right)^\dagger \left( D_M X \right) + V_X( |X| ) \right] + \frac{1}{4} F_{MN} F^{MN} \right\},
	\label{eq:5}
\end{equation}
where \( g \) is the metric determinant, \( \phi(z) \) is the dilaton field, the covariant derivative is defined as \( D_M X = \partial_M X - i A_M X \) and the field strength is expressed as \( F_{MN} = \partial_M A_N - \partial_N A_M \). This study focuses on the scalar background and the \( U(1) \) current, so the \( SU(2) \) gauge fields in the original soft-wall model are neglected. For the \( U(1) \) gauge field \( A_M \) corresponding to the current operator \( O_\mu = \langle q \gamma_\mu q \rangle \), only the time component \( A_t \) and the angular component \( A_\theta \) are retained.These two components describe the local charge density and the angular current \( O_\theta = \langle q \gamma_\theta q \rangle \), respectively, where the latter is necessarily nonzero in a rotating system.

Following existing studies ~\cite{Chelabi:2015gpc,Li:2016gfn,Chelabi:2015cwn}, we adopt a specific form of the dilaton field to describe chiral phase transitions as
\begin{equation}
	\phi(z) = -\mu_1 z^2 + \left( \mu_1 + \mu_0 \right) z^2 \tanh\left(\mu_2 z^2\right),
	\label{eq:6}
\end{equation}
where the parameters are determined by fitting the Regge slope of light mesons, the vacuum expectation value of the condensate, and the chiral phase transition temperature. Their numerical values are set to \(\mu_0 = (0.43 \, \text{GeV})^2\), \(\mu_1 = (0.83 \, \text{GeV})^2\), and \(\mu_2 = (0.176 \, \text{GeV})^2\) ~\cite{Chelabi:2015gpc,Li:2016gfn,Chelabi:2015cwn}.

The spontaneous breaking of \( SU(N_f)_L \times SU(N_f)_R \) symmetry to its \( SU(N_f)_V \) subgroup is realized by a nonzero vacuum expectation value of the complex scalar field \( X \), which takes the form \( X_0 = \frac{\chi}{\sqrt{2N_f}} I_{N_f \times N_f} \), where \( N_f \) is the number of quark flavors and \( I \) is the identity matrix. This study focuses on the case \( N_f = 2 \) (two-flavor quarks). According to the AdS/CFT correspondence, the asymptotic behavior of the scalar field \( \chi(z) \) near the boundary (\( z \to 0 \)) is
\begin{equation}
	\chi(z) = m_q \zeta z + \cdots + \frac{\sigma}{\zeta} z^3,
	\label{eq:7}
\end{equation}
where \(m_q\) is the quark mass, \(\sigma\) is the quark condensate, and \(\zeta = \frac{\sqrt{3}}{2\pi}\) is a normalization constant. In the chiral limit (\(m_q = 0\)), the potential term \(V(\chi)\) determines the order of the phase transition (first-order or second-order). We adopt the following form of the potential and choose appropriate parameters to realize the corresponding phase transition:
\begin{equation}
	V(\chi) \equiv \operatorname{Tr} \left[ V_X (|X|) \right] = \frac{m_5^2}{2} \chi^2 + v_3 \chi^3 + v_4 \chi^4,
	\label{eq:8}
\end{equation}
where the five-dimensional mass squared of the scalar field is set to \(m_5^2 = -3\) (satisfying the requirements of the holographic principle). When the cubic coupling \(v_3\) is nonzero, it corresponds to first-order phase transition, and the quartic coupling \(v_4\) determines the value of the quark condensate at zero temperature. Following existing studies ~\cite{Chelabi:2015gpc,Li:2016gfn,Chelabi:2015cwn}, we select two sets of parameters: \((v_3, v_4) = (0, 8)\) and \((v_3, v_4) = (-3, 8)\), corresponding to second-order and first-order phase transitions in the chiral limit, respectively.

In this work, we do not attempt to model the entire $\mu$-dependent phase structure (e.g., a first-order transition at large $\mu$). Instead, we focus on the combined effects of rotation and a moderate, finite $\mu$ within the probe approximation, which is sufficient for studying the additive suppression of the chiral condensate and the shift of the phase boundary. It should be noted that the minimal probe approximation model adopted in this paper, which utilizes a fixed AdS-RN metric and a quartic potential function, cannot simultaneously produce crossover phenomena at small chemical potential $\mu$ and a first-order phase transition at large $\mu$ in the absence of rotational effects. To obtain such a complete phase diagram, a fully backreacted dynamical metric is indeed required, combined with a more complex dilaton potential function (see references ~\cite{Li:2011hp,Cai:2012xh,He:2013qq} below).

For a rotating system with fixed \( \Omega \), the scalar field \( \chi \) and gauge field \( A_M \) depend only on the fifth-dimensional coordinate \( z \) and the radial coordinate \( r \); they do not depend on the polar angle \( \theta \) or the vertical coordinate \( x_3 \).

Therefore, the scalar field \(\chi\) and the gauge field \(A_M\) are given as
\begin{equation}
	\chi = \chi(z, r), \quad A_M = A_M(z, r).
	\label{eq:9}
\end{equation}

We consider a rotating system that is stable and time-independent. In such a rotating system, one might expect rotational symmetry in the \(\theta\) direction. Therefore, we assume that both the scalar and gauge fields do not depend on the polar angle \(\theta\). In addition, for simplicity, we take the length of the system in the \(x_3\) direction to be infinite and assume that all fields are homogeneous in this direction (of course, for a real system this is not true, and we will leave a thorough analysis for future work). In the \(A_z = 0\) gauge, \(A_\theta\) is the only nonzero component of the gauge field and is dual to the polar current operator \(\langle \bar{q} \gamma^\theta q \rangle\). To preserve conservation of the vector current \(j^\mu = \bar{q} \gamma^\mu q\) of the system, we set the radial current \(\bar{q} \gamma^r q = 0\) and the axial current \(\bar{q} \gamma^{x_3} q = 0\), i.e., the system is in a steady state with no radial or vertical flows. The nonzero polar component \(A_\theta\) corresponds, in the dual field theory, to an effective polarization term \(\vec{\Omega} \cdot \vec{J}\), where \(\vec{J}\) is the angular momentum. Therefore, the equations of motion derived from the action in Eq.~\eqref{eq:5} are
\begin{equation}
	\begin{aligned}
		& \partial_z\left(f\partial_z\chi\right)-f\left(\frac{1}{z}+\phi'\right)\partial_z\chi+\partial_r^2\chi+\frac{\partial_r\chi}{r} \\
		& \quad + \left(\frac{3}{z^2}+\frac{f'}{z}-\frac{3f}{z^2}-\frac{f\phi'}{z}-\frac{3v_3\chi}{z}-4v_4\chi^2-\frac{A_\theta^2}{r^2}\right)\chi = 0,
	\end{aligned}
	\label{eq:10}
\end{equation}
\begin{equation}
	\partial_z \left( f \partial_z A_\theta \right) - f \left( \frac{1}{z} + \phi' \right) \partial_z A_\theta + \partial_r^2 A_\theta - \frac{\partial_r A_\theta}{r} - \chi^2 A_\theta = 0,
	\label{eq:11}
\end{equation}
where \( \chi \) is substituted by \(z \chi \). Following existing research ~\cite{Domenech:2010nf,Keranen:2009ss}, we adopt specific boundary conditions for solving the equations of motion at the AdS conformal boundary as
\begin{equation}
	\chi|_{z=0} = m_q \zeta, \quad A_\theta|_{z=0} = \Omega(r) r^2.
	\label{eq:12}
\end{equation}

For causality considerations, the polar direction is restricted to a finite radius \(R\) (i.e., the system has a finite size). At the edge \(r = R\), we impose boundary conditions on \(\chi\) and \(A_{\theta}\). For \(\chi\), we consider either the Neumann condition \(\partial_r \chi |_{r=R} = 0\) or the Dirichlet condition \(\chi|_{r=R} = 0\). For \(A_{\theta}\), we consistently impose the condition \(A_{\theta} |_{r=R} = \Omega(r) R^2\) to ensure finite angular momentum current at the boundary. This choice is common in holographic studies of rotation ~\cite{Domenech:2010nf}.

We impose either Neumann (\(\partial_r \chi|_{r=R} = 0\)) or Dirichlet (\(\chi|_{r=R} = 0\)) condition at \(r = R\). Neumann corresponds to a free boundary where the condensate remains constant near the edge. Dirichlet mimics a rigid boundary that forces chiral symmetry restoration. We study both to assess sensitivity.

Physically, the Neumann condition $\partial_r \chi|_{r=R} = 0$ corresponds to a free boundary where the chiral condensate can naturally taper off without a hard wall, suitable for a diffuse QGP fireball. The Dirichlet condition ~$\chi|_{r=R} = 0$  ~mimics a rigid confinement boundary that forces chiral symmetry restoration at the edge. For typical central and semi-central heavy-ion collisions, the QGP boundary is not perfectly rigid, so Neumann is more physically motivated. However, we present both to quantify the uncertainty related to finite-size effects and to facilitate comparison with idealized models.

At the center, to ensure the smoothness of the field distribution, polar central boundary conditions are set for the scalar field and the gauge field.
\begin{equation}
	\partial_r \chi |_{r=0} = 0, \quad A_\theta |_{r=0} = 0.
	\label{eq:bc_r0}
\end{equation}

We note that our ansatz assumes a stationary, rigidly rotating system with no radial flow ($A_{r}$) and no dependence on $\theta$ or $x_{3}$. In a realistic finite-volume rotating plasma, the spatially varying chiral condensate may generate radial pressure gradients, potentially inducing a nonzero $A_{r}$. Our neglect of $A_{r}$ is thus an approximation valid for slow rotation and small radial gradients. A full treatment would require solving coupled equations for $A_{r}$ and $A_{\theta}$, which is beyond the scope of this work. Nevertheless, our model captures the leading-order effect of rotation via the $\Omega\cdot J$  coupling, as in previous studies ~\cite{Chelabi:2015gpc,Li:2016gfn,Chelabi:2015cwn, Domenech:2010nf}.

The calculation process for the chiral condensate \(\sigma\) is summarized as follows: First, the equations of motion for the scalar field \(\chi\) and the \(U(1)\) gauge field $A_\theta$ are obtained from the action in Eq.~\eqref{eq:5} and the relevant settings. Then, the boundary conditions are imposed on the equations of motion to solve for the scalar field \(\chi\). Finally, the chiral condensate \(\sigma\) is obtained according to Eq.~\eqref{eq:7}.

\section{The influence of temperature, chemical potential, and rotation on chiral condensation}\label{sec:3}

In this section, we will investigate the impact of temperature-chemical potential rotation on the spatial distribution of chiral condensation. The distribution of angular velocity in space can affect the shape of chiral condensation. For simplicity, we assume \(\Omega = \text{constant}\). Causality requires the system to be confined within a finite radius. We fix the radius at \(4\ \text{fm}\) (approximately \(20\ \text{GeV}^{-1}\)), typical for QGP. To respect causality, the maximum angular velocity is limited to \(0.05\ ~\text{GeV}\).

In this section, we investigate the distribution of chiral condensate in radial coordinate \(r\). Since whether the phase transition is first-order, second-order, or crossover is not important for qualitative results, this example is illustrated using the potential in Eq.~\eqref{eq:8} with parameters \((m_q, v_3, v_4) = (0, -3, 8)\).

\begin{figure}
	\centering
	\includegraphics[width=0.8\linewidth]{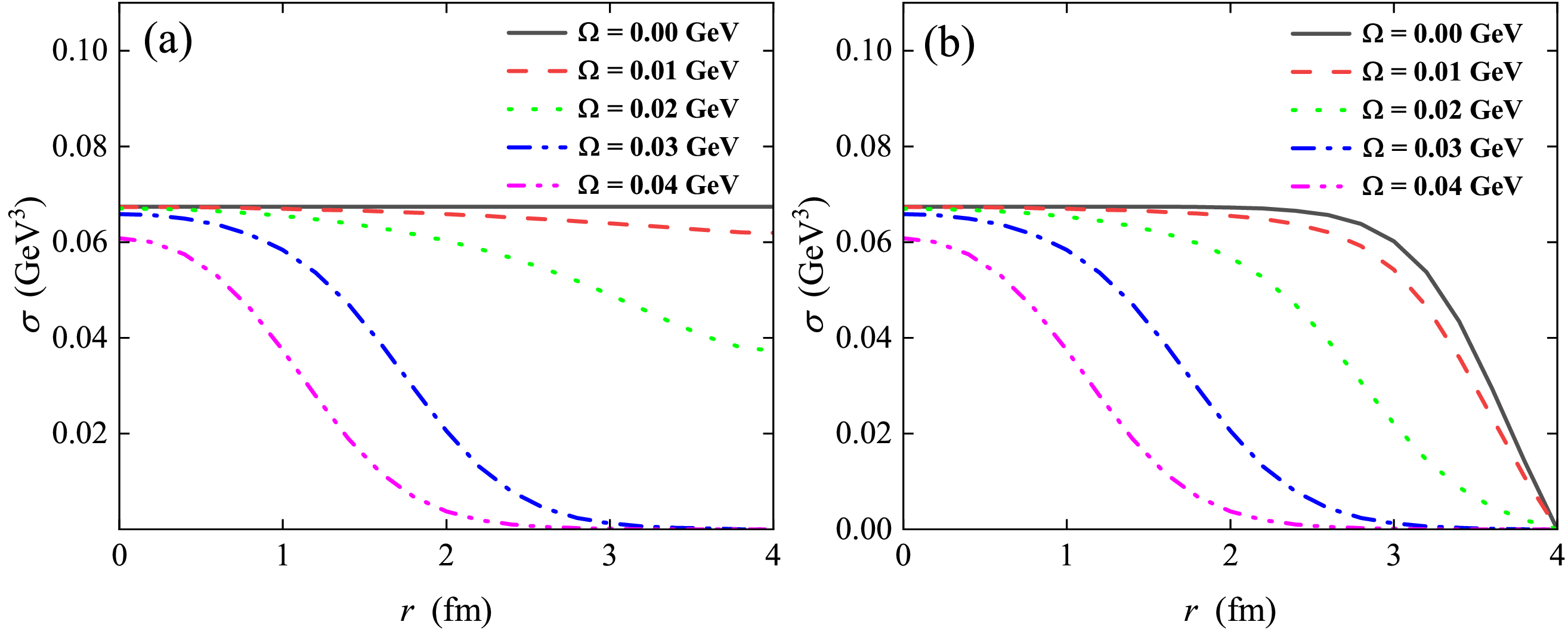}
	\caption{The radial distribution of the chiral condensate \(\sigma\) for different angular velocities \(\Omega\) at fixed temperature \(T = 0.16 \, \text{GeV}\) and chemical potential \(\mu = 0.15 \, \text{GeV}\). Panel (a) corresponds to Neumann boundary conditions, and panel (b) to Dirichlet boundary conditions.}
	\label{fig:1}
\end{figure}

\begin{figure}
	\centering
	\includegraphics[width=0.8\linewidth]{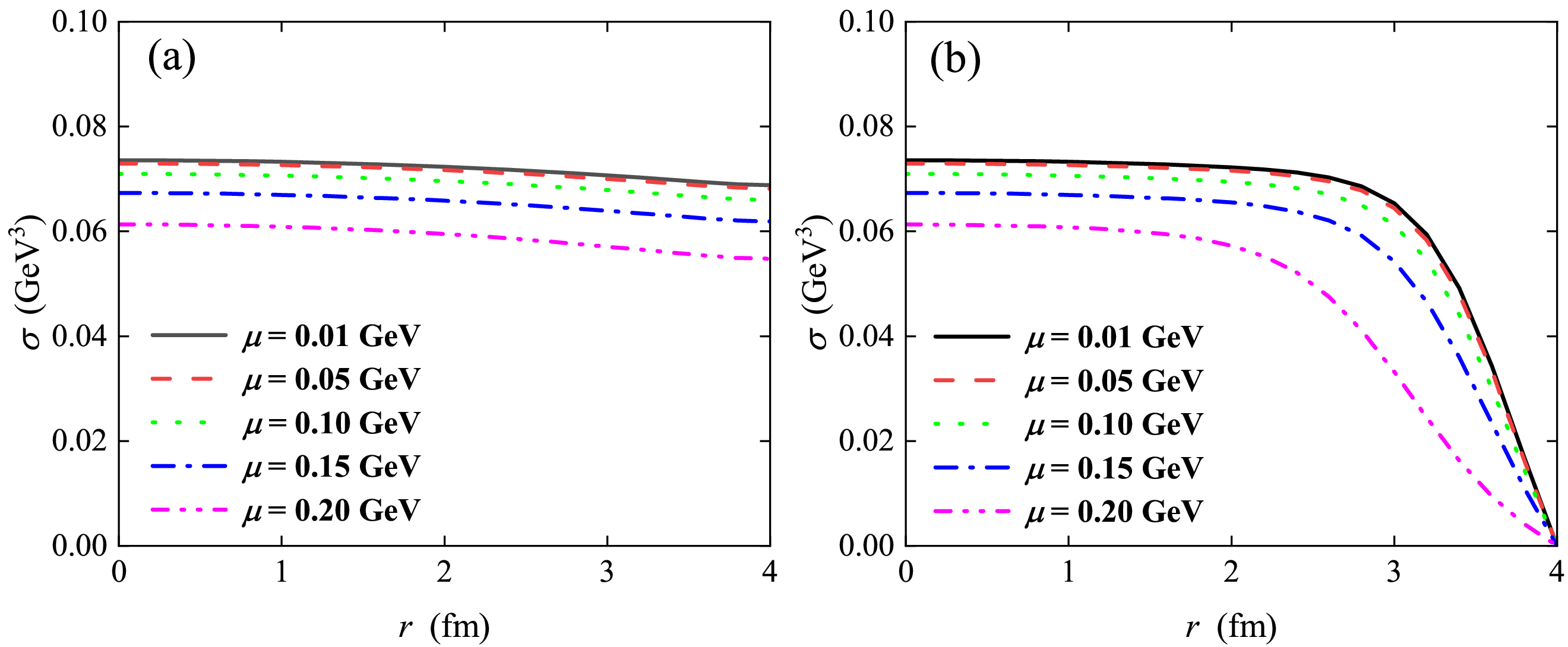}
	\caption{The radial distribution of the chiral condensate \(\sigma\) for different chemical potentials \(\mu\) at fixed temperature \(T = 0.16 \, \text{GeV}\) and angular velocity \(\Omega = 0.01 \, \text{GeV}\). Panel (a) corresponds to Neumann boundary conditions, and panel (b) to Dirichlet boundary conditions.}
	\label{fig:2}
\end{figure}

\begin{figure}
	\centering
	\includegraphics[width=0.8\linewidth]{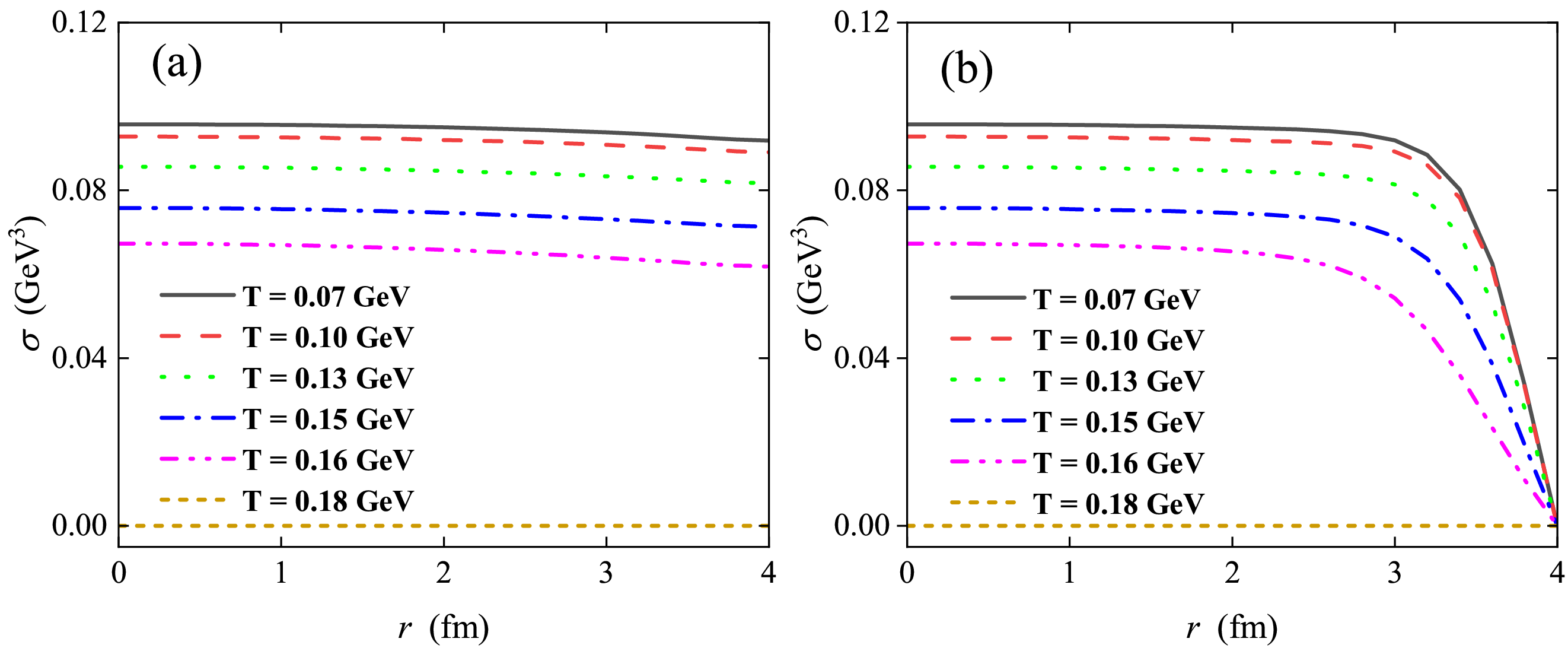}
	\caption{The radial distribution of the chiral condensate \(\sigma\) as a function of the radial coordinate \(r\) at fixed chemical potential \(\mu = 0.15 \, \text{GeV}\) and angular velocity \(\Omega = 0.01 \, \text{GeV}\), for different temperatures. Panel (a) corresponds to Neumann boundary conditions, and panel (b) to Dirichlet boundary conditions.}
	\label{fig:3}
\end{figure}

Figures 1--3 show the radial distribution of the chiral condensate \(\sigma(r)\) under different conditions. Figure~\ref{fig:1} fixes \(T = 0.16 \, \text{GeV}, \, \mu = 0.15 \, \text{GeV}\) and varies \(\Omega\): as \(\Omega\) increases, \(\sigma(r)\) becomes more suppressed near the edge, vanishing there for sufficiently large \(\Omega\) (left: Neumann; right: Dirichlet). We define the critical angular velocity $\Omega_c$ as the smallest $\Omega$ ~for which the chiral condensate at the boundary $r=R$ becomes zero (within numerical precision). Under Neumann conditions, $\Omega_c \approx 0.023$ GeV, while under Dirichlet conditions, $\Omega_c \approx 0.021$ GeV for  $T=0.16$ GeV, $\mu=0.15$ GeV. Figure~\ref{fig:2} fixes \(T = 0.16 \, \text{GeV}, \, \Omega = 0.01 \, \text{GeV}\) and varies \(\mu\): higher \(\mu\) globally reduces \(\sigma\) but does not alter the spatial pattern. Fig.~\ref{fig:3} fixes \(\mu = 0.15 \, \text{GeV}, \, \Omega = 0.01 \, \text{GeV}\) and varies \(T\): increasing \(T\) suppresses \(\sigma\) everywhere, with the edge vanishing first, indicating a position-dependent critical temperature.Unlike the non-rotating case (not shown), where \(\sigma(r)\) is uniform, the finite \(\Omega\) in Fig.~3 creates a radial gradient.

The results in Fig.~\ref{fig:1} demonstrate that rotation induces a spatially dependent chiral phase transition. Even when the system is globally at a uniform temperature, the effective critical temperature for chiral symmetry restoration becomes position-dependent: regions farther from the rotation axis experience a lower effective transition temperature. This leads to a situation where the edge of the system can be in the chiral-symmetric phase while the center remains in the chiral-broken phase. Such inhomogeneous behavior is a key signature of rotational effects in QCD matter and has important implications for understanding the phase structure of quark-gluon plasma produced in non-central heavy-ion collisions, where large angular velocities are present.

The results in Fig.~\ref{fig:2} indicate that the chemical potential \(\mu\) reduces the magnitude of \(\sigma\) uniformly across \(r\) induced by rotation, leaving the normalized radial profile \(\sigma(r)/\sigma(0)\) almost unchanged. This suggests that while baryon density weakens chiral symmetry breaking across the system, the rotational effect remains the dominant mechanism driving spatial variation. The persistence of the central plateau under Dirichlet boundary conditions further highlights the sensitivity of the condensate profile to boundary choices, which has implications for modeling finite-size systems such as the quark-gluon plasma produced in heavy-ion collisions. Additionally, the fact that the condensate eventually vanishes uniformly at sufficiently high temperature (as seen in related figures) reinforces that temperature and chemical potential both drive chiral restoration, but their interplay with rotation leads to a rich, spatially dependent phase structure.

The physical significance of Fig.~\ref{fig:3} lies in revealing that rotation breaks the spatial uniformity of the chiral phase transition. The observation that the chiral condensate vanishes first at the edge as temperature increases implies that different radial regions of the system undergo chiral restoration at different temperatures. Such spatial inhomogeneity has important implications for interpreting experimental signatures from non-central heavy-ion collisions, where large angular velocities are generated. Additionally, the sensitivity to boundary conditions highlights the role of finite-size effects and confinement geometry, which are crucial for realistic modeling of the QGP fireball.
\section{The dependence of chiral phase diagram on chemical potential and rotation}\label{sec:4}
It is found that the normalized radial profile \(\sigma(r)/\sigma(0)\) remains almost unchanged because \(\mu\) reduces the magnitude of \(\sigma\) uniformly across \(r\) induced by rotation. However, because the phase transition temperature depends on \(\mu\), the overall phase boundary in the \(T\)-\(\Omega\) plane is shifted. Including \(\mu\) makes the phase diagram more realistic for the rotating QGP produced in heavy-ion collisions, where both rotation and finite baryon density coexist. We use the chiral condensate at the center (\(r = 0\)) as the order parameter, since it reaches its maximum there and the chemical potential mainly suppresses its magnitude without changing the spatial pattern induced by rotation. Using this order parameter, we construct the \(T\)-\(\Omega\) phase diagrams for different \(\mu\). We can also compute the \(T\)-\(\mu\) phase diagram for different \(\Omega\) at \(r = 0\).
\begin{figure}
	\centering
	\includegraphics[width=0.6\linewidth]{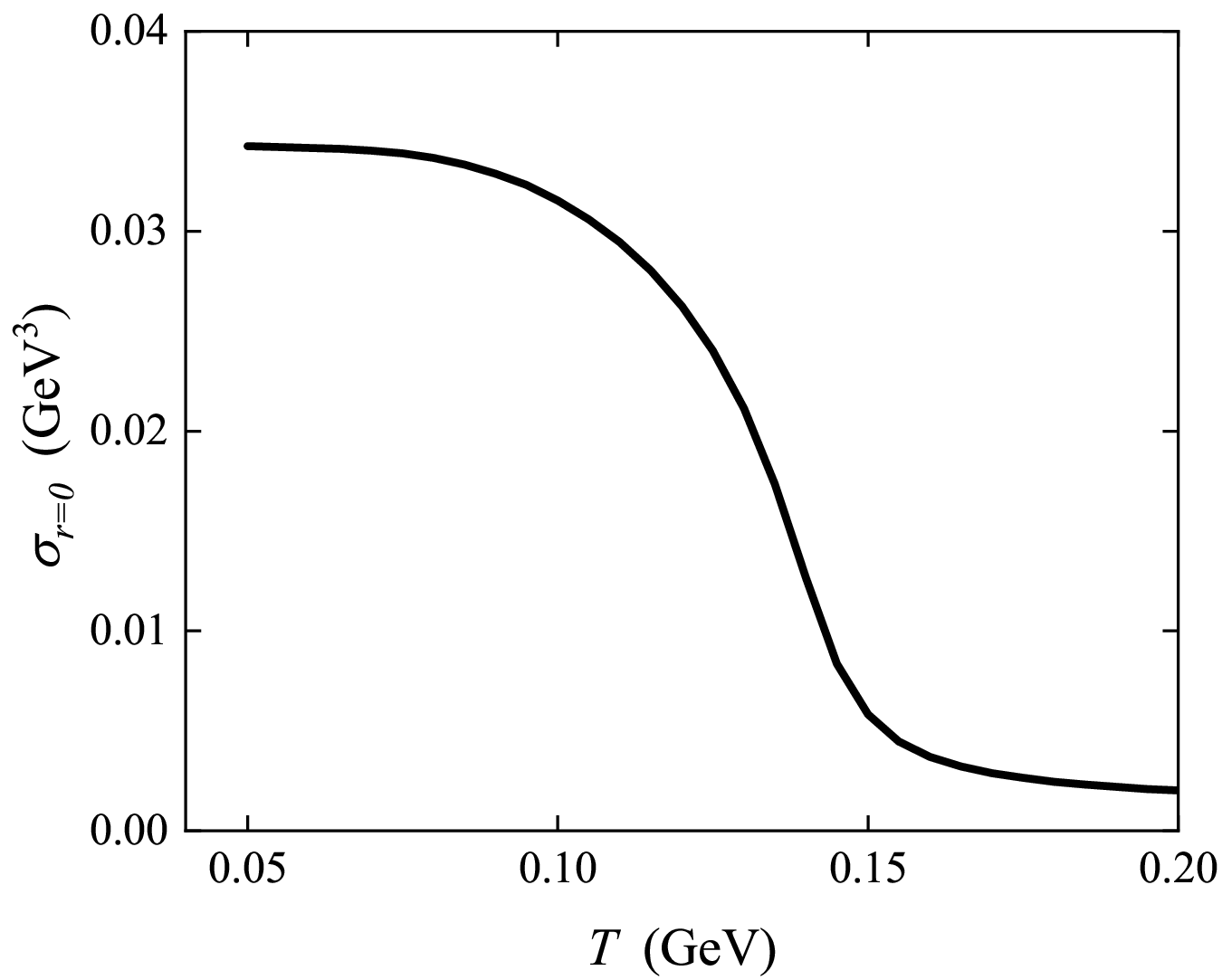}
	\caption{Temperature dependence of the chiral condensate at the center (\(r = 0\)) under Neumann boundary conditions, with \(\Omega = 0.01 \, \text{GeV}\) and \(\mu = 0.1 \, \text{GeV}\). The condensate decreases slowly at low temperatures and drops sharply near the phase transition. The point of the steepest slope (maximum of \(|\partial \sigma/\partial T|\)) defines the pseudocritical temperature \(T_c\).}
	\label{fig:4}
\end{figure}

As shown in Fig.~\ref{fig:4}, the central chiral condensate \(\sigma\) remains large at low temperatures, then decreases slowly, and finally drops rapidly in a narrow temperature interval. This behavior is typical for a crossover or a second-order phase transition. We define the pseudocritical temperature \(T_c\) as the point where the slope \(\frac{d\sigma}{dT}\) reaches its maximum. We extract \(T_c\) from the inflection point of \(\sigma(T)\) (see Fig.~\ref{fig:4}) and repeat for each \(\Omega\) and \(\mu\) to obtain the phase boundaries in Figs.~5--7.

\begin{figure}
	\centering
	\includegraphics[width=0.8\linewidth]{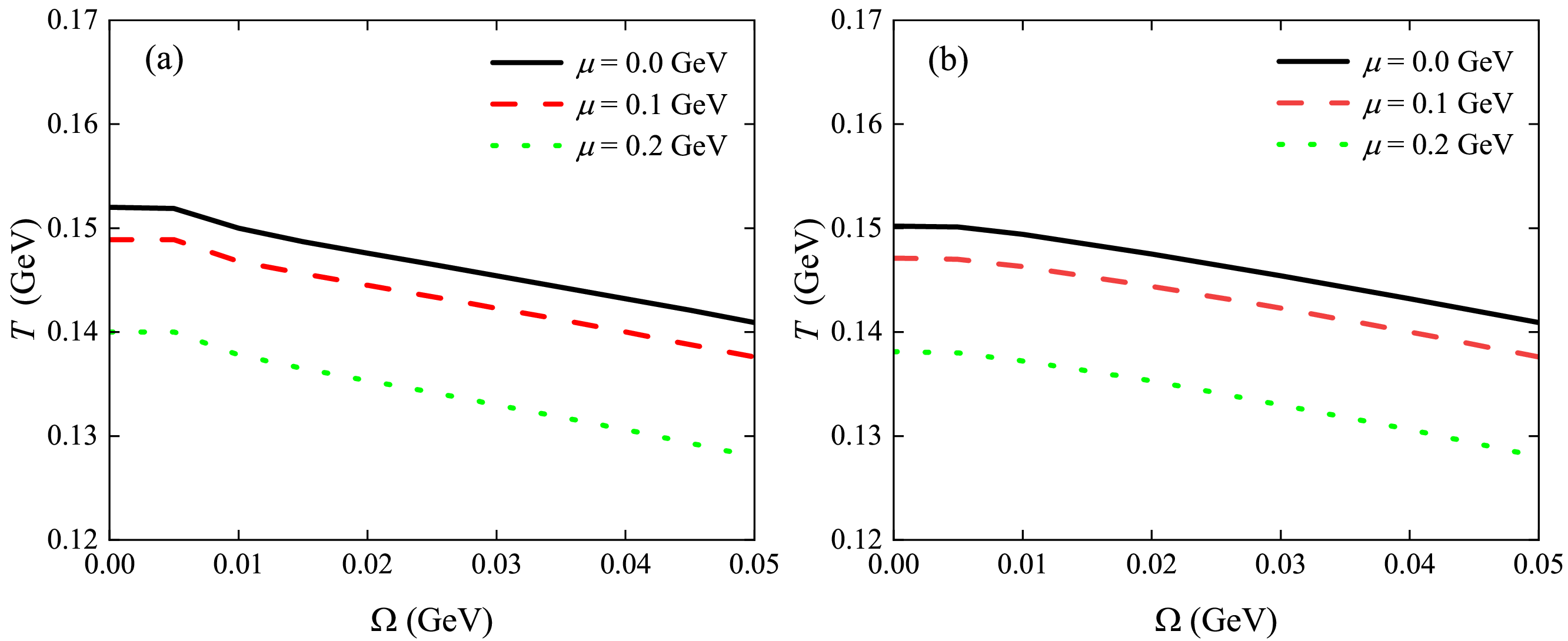}
	\caption{Temperature-angular velocity (\(T\)-\(\Omega\)) phase diagrams for chemical potentials \(\mu = 0\), \(0.1\), and \(0.2\) GeV. The phase transition temperature is determined by the chiral condensate at the center of the system. Panel (a) is under Neumann boundary conditions, and panel (b) is under Dirichlet boundary conditions.}
	\label{fig:5}
\end{figure}

Figure ~\ref{fig:5} presents the \(T\)-\(\Omega\) phase diagrams for three different chemical potentials under two types of boundary conditions (Neumann and Dirichlet). The phase boundary is defined using the chiral condensate at the center (\(r = 0\)) as the order parameter. Figure~\ref{fig:5} illustrates how the critical temperature for chiral symmetry restoration depends on both the angular velocity \(\Omega\) and the chemical potential \(\mu\). It shows that as \(\Omega\) increases, the critical temperature decreases monotonically. Moreover, for a fixed \(\Omega\), increasing \(\mu\) lowers the phase transition temperature. A comparison between the two panels reveals that the critical temperatures under Neumann boundary conditions are systematically higher than those under Dirichlet boundary conditions, though the overall trend of the phase boundary remains similar.

The physical significance of Fig.~\ref{fig:5} lies in its demonstration that both rotation and baryon density suppress chiral symmetry restoration, but they do so in a way that the suppression effects are additive. The downward shift of the phase boundary with increasing \(\mu\) confirms that a finite chemical potential, like rotation, acts as a catalyst for chiral restoration. This result is particularly relevant for understanding the QCD phase structure under conditions similar to those in non-central heavy-ion collisions, where both large angular momentum and finite baryon density are present. The systematic difference between Neumann and Dirichlet boundary conditions highlights that the choice of boundary condition, which reflects the physical confinement geometry of the system, can quantitatively affect the phase transition temperature. This sensitivity underscores the importance of properly modeling finite-size and edge effects in holographic studies of rotating QCD matter.
\begin{figure}
	\centering
	\includegraphics[width=0.8\linewidth]{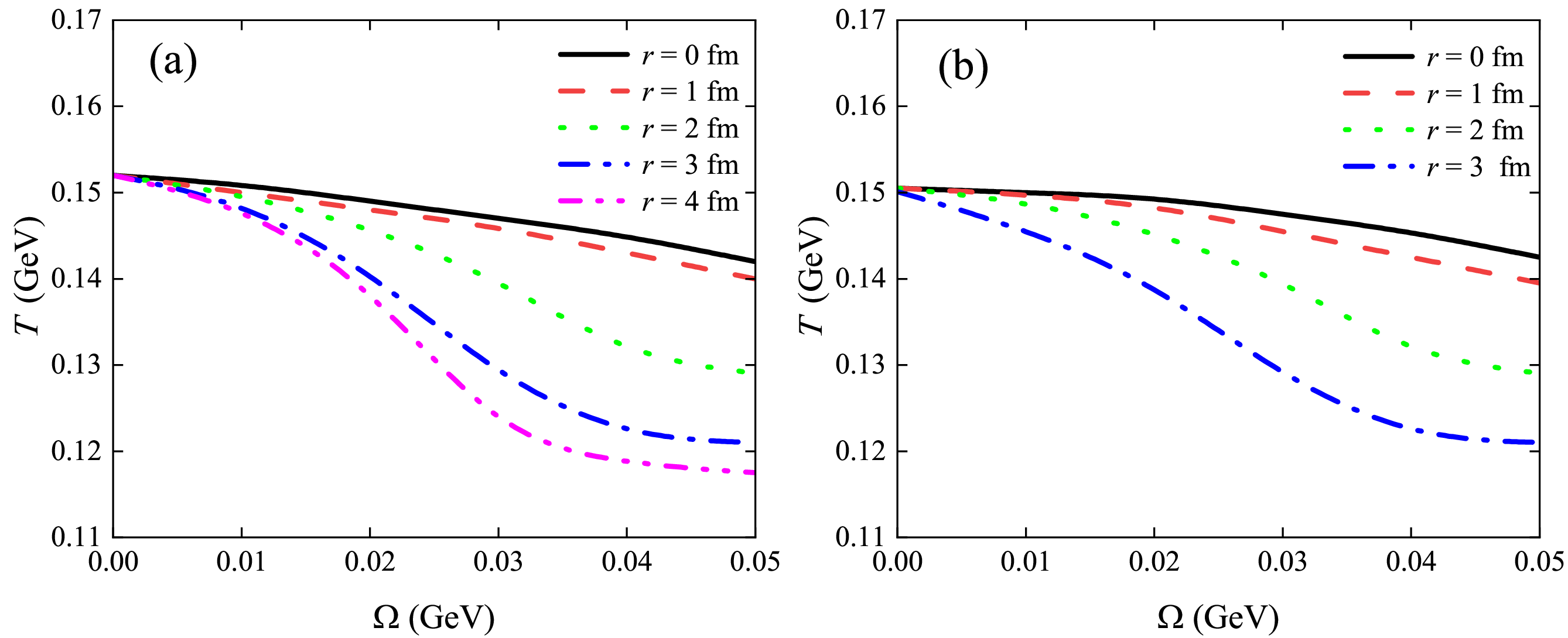}
	\caption{Temperature-angular velocity (\(T\)-\(\Omega\)) phase diagrams at different radial distances \(r\) from the rotation axis, with chemical potential \(\mu = 0\). Panel (a) corresponds to Neumann boundary conditions, and panel (b) to Dirichlet boundary conditions. The phase transition temperatures are extracted from the chiral condensate at \(r = 0, 1, 2, 3\), and \(4\) fm. Under Dirichlet boundary conditions, the \(T\)-\(\Omega\) curve at \(r = 4 \, \text{fm}\) is not shown because the method for determining the phase transition temperature becomes invalid due to non-monotonic behavior.}
	\label{fig:6}
\end{figure}

Figure~\ref{fig:6} investigates the spatial dependence of the chiral phase transition in a rotating system by presenting \(T\)-\(\Omega\) phase diagrams for different radial distances \(r\) from the rotation axis. The results show that when \(\Omega = 0\), the phase transition temperature is the same at all radial positions, as expected for a non-rotating system. Once rotation is turned on, the critical temperature becomes position-dependent: regions farther from the rotation axis exhibit lower phase transition temperatures. This effect becomes more pronounced as the angular velocity increases. The two boundary conditions produce similar qualitative behavior, though quantitative differences exist, and under Dirichlet conditions the determination of the phase transition temperature at the largest \(r\) becomes problematic due to the condensate profile's shape. This confirms that rotation is the sole source of the radial dependence.

The physical significance of Fig.~\ref{fig:6} is that it directly demonstrates how rotation breaks the spatial uniformity of the chiral phase transition. In a rotating system, chiral symmetry restoration does not occur simultaneously throughout the volume; instead, the periphery of the system restores chiral symmetry at a lower temperature than the center. This radial dependence of the critical temperature implies that in a realistic rotating quark-gluon plasma, there can coexist phases with broken and restored chiral symmetry at the same global temperature. Such a configuration may lead to novel experimental signatures, such as position-dependent particle production or anisotropic flow patterns. Furthermore, the observation that the phase transition temperature drops more rapidly with \(\Omega\) at larger \(r\) highlights that the rotational suppression of chiral condensation is strongest near the boundary, consistent with the earlier observation that the condensate vanishes at the edge under sufficiently high rotation. This radial inhomogeneity is a key feature of rotating strong-interaction matter and distinguishes it from systems under uniform external conditions.
\begin{figure}
	\centering
	\includegraphics[width=0.8\linewidth]{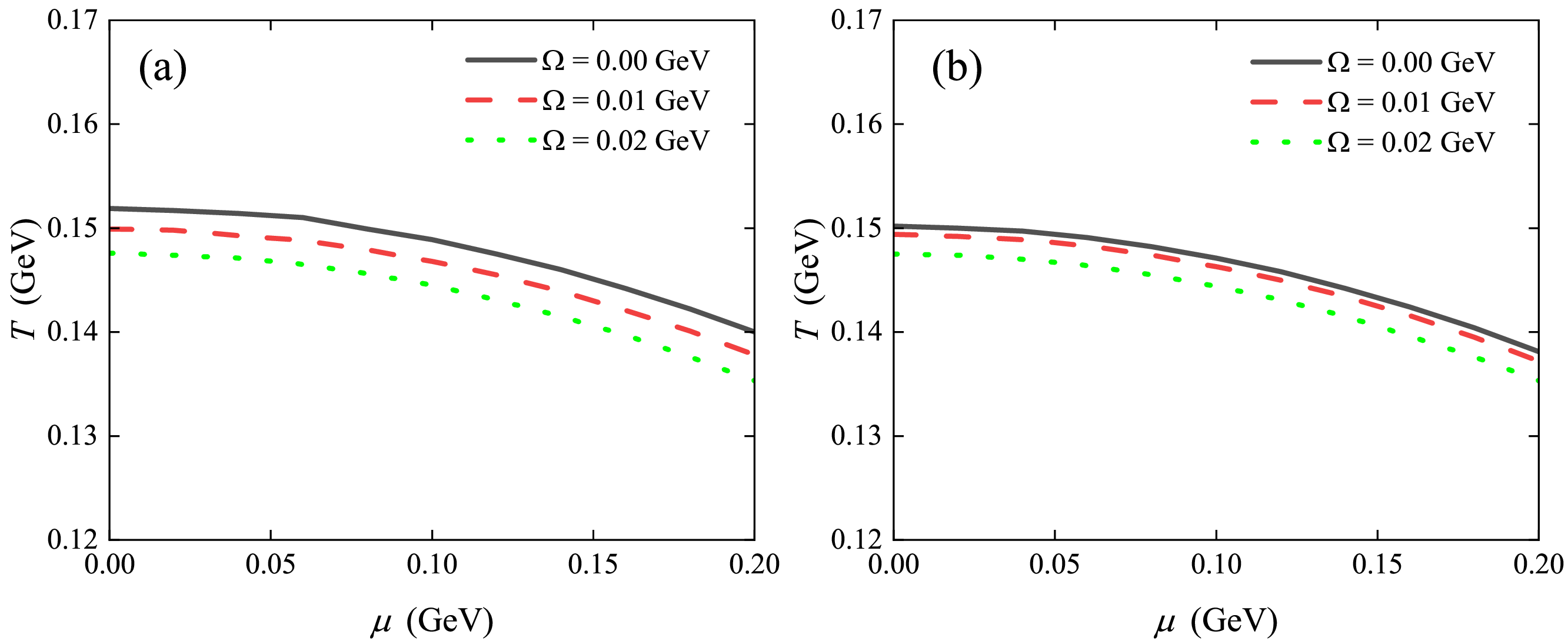}
	\caption{Temperature-chemical potential (\(T\)-\(\mu\)) phase diagrams at different angular velocities \(\Omega\), evaluated at the center \(r = 0\). Panel (a) is for Neumann boundary conditions, and panel (b) for Dirichlet boundary conditions.}
	\label{fig:7}
\end{figure}

Figure~\ref{fig:7} illustrates how the chiral phase transition temperature \(T_c\) depends on the quark chemical potential \(\mu\) for different fixed angular velocities \(\Omega\), under two types of boundary conditions (Neumann and Dirichlet). It shows that increasing \(\mu\) lowers the critical temperature, and this suppression effect is enhanced by larger \(\Omega\). The phase boundaries shift downward systematically as either \(\mu\) or \(\Omega\) increases.

The results demonstrate that both the chemical potential and rotation act as catalysts for chiral symmetry restoration. Figure 7 further quantifies the additive suppression: higher \(\Omega\) enhances the sensitivity of \(T_c\) to \(\mu\), and vice versa. This implies that in realistic non-central heavy-ion collisions, where both large angular momentum and finite baryon density coexist, the chiral phase transition occurs at lower temperatures than in non-rotating or zero-density systems. Furthermore, the comparison between Neumann and Dirichlet boundary conditions reveals that finite-size and edge effects quantitatively affect the phase boundary, highlighting the importance of modeling the confinement geometry correctly in holographic studies of rotating QCD matter.

\section{SUMMARY AND CONCLUSIONS}\label{sec:5}
Comparing with Ref.~\cite{Chen:2022mhf}, our work confirms their findings at $\mu = 0$ and extends them to finite  $\mu$. While Ref.~\cite{Chen:2022mhf} focused on finite-size effects and boundary conditions without  $\mu$, we find that introducing  $\mu$ preserves the qualitative spatial patterns but uniformly reduces the condensate magnitude. Furthermore, the decrease of $\Omega_{c}$ with increasing $\mu$ is a new observation not reported previously. In this work, we have studied the combined effects of rotation and finite quark chemical potential on the chiral phase transition within the soft-wall holographic QCD model. Our main findings are:

(1) Rotation induces spatial inhomogeneity - at fixed \(T\) and \(\mu\), the suppression of the chiral condensate by \(\Omega\) is stronger at the edge than at the center. Beyond a critical \(\Omega\), the condensate vanishes at the boundary while remaining nonzero at the center.

(2) Chemical potential acts as a uniform suppressor - raising \(\mu\) reduces the overall magnitude of the condensate but does not alter the spatial pattern induced by rotation.

(3) Additive lowering of \(T_c\) - both \(\Omega\) and \(\mu\) lower the chiral phase transition temperature, and their suppression effects are additive. In a rotating system, \(T_c\) becomes position-dependent: it decreases with distance from the rotation axis.

(4) Boundary-condition dependence - Neumann and Dirichlet boundary conditions yield similar qualitative behavior but differ quantitatively, with Dirichlet conditions causing a steeper drop of the condensate near the edge.

These results reveal a rich, spatially dependent phase structure in rotating QCD matter, relevant for non-central heavy-ion collisions where both large angular momentum and finite baryon density coexist. Future work may include full backreaction, realistic \(\Omega\) profiles, and vortex formation.

	\section*{Acknowledgments}
	The authors thank Yidian Chen for the valuable discussions. This work was supported by the National Natural Science Foundation of China (Grants No. 12575144, and No. 11875178).
	
	\section*{References}
	
	\nocite{*}
	\bibliography{long}

\end{document}